\documentstyle{mn}
\input{epsf}
\newcommand{\be}{\begin{eqnarray}}
\newcommand{\ee}{\end{eqnarray}}

\title[Characteristics of metal systems ]
{Statistical characteristics of the observed metal systems
and problems of reionization}
\author[Demia\'nski \&  Doroshkevich ]
       {M. Demia\'nski$^{1,2}$,  A.G. Doroshkevich$^{3}$,\\
        $1$Institute of Theoretical Physics,
                       University of Warsaw,
                       00-681 Warsaw, Poland\\
        $2$Department of Astronomy, Williams College,
           Williamstown, MA 01267, USA\\
        $3$Astro Space Center of Lebedev Physical
           Institute of  Russian Academy of Sciences,
                        117997 Moscow,  Russia\\
}

\date{Accepted ...,
      Received ...,
        in original form ... .}

\begin{document}
\maketitle

\begin{abstract}
We analyze basic properties of about 200 metal systems observed
in high resolution spectra of 9 quasars by Boksenberg,
Sargent\,\&\, Rauch (2003). The measured Doppler parameters for
the hydrogen and carbon lines are found to be different by a
factor $\sim\sqrt{m_C/ m_H}$ what indicates the domination of
the thermal broadening and high degree of relaxation for
majority of the observed metal systems. The analysis of the mean
separation of the metal systems confirms that they can be
located in regions with the typical proper size $R_{cl}\sim 0.1
- 0.25h^{-1}$ Mpc what corresponds to the typical sizes of
protogalaxies with the baryonic masses $M_b\approx
10^9-10^{10}M_\odot$. The metal abundances observed at redshifts
$z=3 - 6$ within both IGM and galaxies are found to be quite
similar to each other what indicates the strong interaction of
young galaxies with the IGM and the important role of dwarf
satellites of host galaxies in such interaction. On the other
hand, this metal abundance can be considered as the integral
measure of nuclear reactions within any stars what in turn
restricts the contribution of the stars to the creation of
ionizing UV background. Available now estimates of these
abundances demonstrate that attempts to explain the reionization
of the Universe as a byproduct of the process of synthesis of
metals only are problematic and a significant contribution of
unobserved and/or non thermal UV sources seems to be required.
\end{abstract}


\section{Introduction}

One of the most promising way to study the earlier stages of the
process of galaxy formation and reionization of the Universe is
the analysis of the Ly--$\alpha$ forest and metal systems
observed in spectra of the farthest quasars. The great potential
of such approach was discussed already by Oort (1981, 1984) just
after Sargent et al. (1980) established the intergalactic nature
of the forest. Indeed, the majority of HI absorbers are
associated with the small scale distribution of intergalactic
matter (IGM) along the line of sight at redshifts $z\geq 2$,
when the IGM is not yet strongly clustered and its observed
characteristics can be more easily interpreted. In turn, the
observed metal systems are evidently associated with galaxies
and they characterize the rate of galaxy formation and their
environment. The available Keck and VLT high resolution
observations of the absorption lines provide a reasonable
database and allow one to apply statistical methods for their
analysis. This analysis can be also supplemented by comparison
of basic properties of metal systems with those available for
galaxies at corresponding redshifts. These problems were
discussed in many papers (see, e.g., Bergeron et al. 1992;
Lanzetta et al. 1995; Tytler 1995; Le Brune et al. 1996) and
more recently in Boksenberg, Sargent \& Rauch (2003); Adelberger
et al. (2005b) and Scannapieco et al. (2006).

The enrichment of the IGM by metals can be related to the
starburst driven outflows (Pettini et al. 2001; Frye, Broadhurst
\& Benitez 2002) from the main galaxy or its dwarf satellites
and to the stripping of these satellites. Both these processes
occur but their efficiency is different. Thus, rare richer
multicomponent systems are usually associated with the periphery
of massive galaxies (see, e.g., Adelberger et al. 2005b;
Songaila 2006). Their complex structure demonstrates that they
were formed by a multi step process that involved both explosive
and stripping metal injections. On the contrary, the commonly
encountered poorer metal systems that are related to extended
low density clouds are more probably linked with the dwarf
satellites of the central galaxy. Detailed discussion of this
problem can be found in Scannapieco et al. (2006).

The comparison of Doppler parameters measured for the HI and metals
in the same absorption systems allows also to compare the influence
of the IGM temperature and the macroscopic motions on the structure
of the clouds, to evaluate the degree of relaxation of compressed
matter and to select the small fraction of clouds with the
domination of supersonic macroscopic motions.

The separate problem is the process of reionization of the
Universe. Observations of the farthest quasars (Becker et al.
2001; Djorgovski et al. 2001; Fan et al. 2004) show that the
reionization of the IGM had just been completed at $z\sim 6-7$.
On the other hand, the measurements of temperature fluctuations
of the CMB by WMAP (Hinshaw et al. 2006) suggest that the
reionization of the IGM has began at least at $z\sim 10 - 12$.
Simulations indicate that the first stars and supernovae can be
formed already at $z\sim 40 - 50$ (Reed et al. 2005) but for the
standard cosmological model at $z\geq 10$ the fraction of matter
accumulated by high density objects (``galaxies'') is negligible
($\leq 10^{-6})$ and therefore reionization earlier than at
$z\geq 10$ requires some exotic sources of the UV radiation such
as, for example, antimatter or unstable particles (see, e.g,
Cohen, De Rujula, Glashow 1998; Bambi \& Dolgov 2007; Freese et
al. 2007).

It is commonly believed that the general features of the
reionization process are already established with a reasonable
reliability (see, e.g., Tumlinson et al. 2004; Madau 2007;
Schaerer 2007). At $z\leq 3.5 - 4$ the observed radiation of
quasars dominates the UV ionizing background, but it provides
not more than 20 -- 30\% of the UV background at $z\geq 5$. Now
it is expected that the UV radiation required for the
reionization can be produced by the joint action of galaxies
dominated by Population III and Population II stars (see, e.g.,
Tumlinson et al. 2004; Schaerer 2007) and by sources of non
thermal radiation such as AGNs, miniquasars or black holes (see,
e.g., Madau\,\& \,Rees 2001; Meiksin 2005). However, the
relative contribution of stars, gamma ray bursts (GRBs) and/or
non thermal sources of UV radiation has not been determined yet.
Now the main discussed problems are focused on indirect
estimates and observational restrictions of efficiency of such
sources (see, e.g., Dijkstra, Haiman \& Loeb 2004; Meiksin 2005;
Choudy et al. 2007) and the analysis of their interaction with
the environment (see, e.g., Iliev, Shapiro \& Raga 2004).

� However we still have not seen neither  the Pop III stars
nor the first galaxies dominated by such stars and, so, the
actual properties of such objects remain unknown. It seems that
all galaxies observed at high redshifts --including both the
Ly--breack galaxies (LBGs) and Ly--$\alpha$ emitters (LAEs) --
are dominated by the Pop II stars, and only small part of LAE
demonstrate some features expected for the first galaxies.
Because of this it can be expected that the domination of Pop
III stars could take place during only a short period of
evolution of the first galaxies and these stars acts as a
trigger for the faster formation of Pop II stars and for the
transformation of the first galaxies into the LAEs and LBGs
(Schneider et al. 2002; Ricotti\,\&\,Ostriker 2004; Smith \,\&\,
Sigurdsson 2007; Tornatore et al. 2007; Karlsson et al. 2007).
� Thus the actual properties of the first galaxies could be
quite similar to the observed properties of the LBGs and LAEs
and the contribution of stars (and particularly of the Pop III
stars) to the production of the ionizing UV background and the
process of reionization can be quite limited (see, e.g.,
Shull\,\&\,Venkatesan 2007) while the contribution of
alternative non thermal sources of UV luminosity can be
essential.

Special problem is the existence of significant population of
LAEs at redshifts $z\leq 5 - 6$. It is evident that the life
time of envelopes of HI surrounding these galaxies is finite
owing to the combine action of the observed outer UV background
and internal sources of UV radiation. For the same reasons the
recent formation or successive reconstruction of such envelops
from the highly ionized IGM seem to be quite problematic. On the
other hand, if LAEs represent special population formed from low
ionized IGM at high redshifts then it is necessary to explain
the unexpectedly slow evolution of survived objects. In this
case a clear redshift variations of the number density of LAEs
and their observed properties should be observed.

Here we analyze the sample of 908 high resolution CIV absorbers
presented in Boksenberg, Sargent \& Rauch (2003) and compare the
measured properties of CIV and HI lines. This approach allows us
to demonstrate that the majority of metal systems are associated
with relaxed clouds. Our indirect estimates of the sizes of
metal systems are consistent with direct measurements of
Adelberger et al. (2005b) and Scannapieco et al. (2006). Our
estimates of the CIV abundance in the IGM are similar to that
observed within LBGs and LAEs, what demonstrates the important
contribution of dwarf galaxies in the reionization. At the same
time, the estimates of the metal abundance show that the
radiation of stars only cannot reionize the Universe and the
contributions of these stars and non thermal sources can be at
least comparable.

In this paper we consider the spatially flat $\Lambda$CDM
model of the Universe with the Hubble parameter and mean
density given by:
\[
H^{2}(z) = H_0^2\Omega_m(1+z)^3[1\Omega_\Lambda/\Omega_m
(1+z)^{-3}]\,,
\]
\be
\langle n_b(z)\rangle = 2.4\cdot 10^{-7}
(1+z)^3(\Omega_bh^2/0.02){\rm cm}^{-3}\,,
\label{basic}
\ee
\[
\langle\rho_b(z)\rangle = {3H_0^2\over 8\pi G}\Omega_b(1+z)^3
\approx 4.8\cdot 10^9(1+z)^3\frac{\Omega_bh^2}{0.02}
\frac{M_\odot}{\rm Mpc^3}\,,
\]
\[
\langle\rho_m(z)\rangle = {3H_0^2\over 8\pi G}\Omega_m(1+z)^3,
\quad H_0=100h\,{\rm km/s/Mpc}\,.
\]
Here $\Omega_m=0.3\,\&\,\Omega_\Lambda=0.7$ are the
dimensionless density of matter and dark energy, $\Omega_b
\approx 0.02$ and $h=0.7$ are the dimensionless mean density of
baryons, and the Hubble constant. For $z\geq 1$ the influence of
$\Lambda$--term in (\ref{basic}) becomes negligible and we will
write the Hubble parameter as 
\be 
H(z)\approx H_0\sqrt{\Omega_m}(1+z)^{3/2}\,. 
\label{hz} 
\ee

\section{ The database}
The present analysis is based on 9 high resolution spectra
listed in Boksenberg, Sargent \& Rough (2003). The full sample 
of metal systems contains 908 CIV lines majority of which 
represent the internal structure of rich metal systems. From 
these spectra we select 193 metal systems at $2\leq z\leq 4.4$. 
For 93 of these systems we have also parameters of the HI 
absorbers.

\begin{figure}
\centering
\epsfxsize=7.5cm
\epsfbox{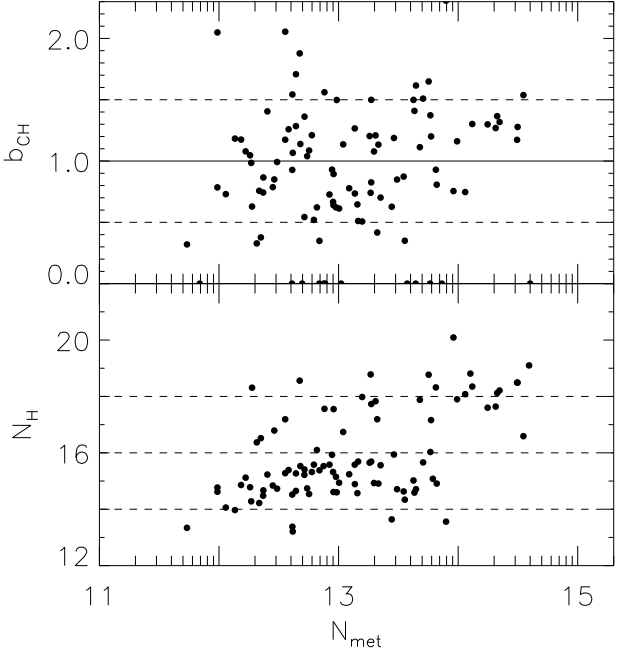}
\vspace{1.cm} 
\caption{For the sample of 93 metal systems with $N_H>0$ functions
$b_{CH}=\sqrt{m_C/m_H}~b_{met}/b_H$, and $N_H$ are plotted vs.
$N_{met}$. } 
\label{ch_bks}
\end{figure}

As is seen from Fig. \ref{ch_bks} the hydrogen and carbon column
densities are weakly correlated and even for richer hydrogen
absorbers with $N_{HI}\geq 10^{16}cm^{-2}$ there are metal
systems with $2\cdot 10^{12}cm^{-2}\leq N_C\leq 6\cdot
10^{14}cm^{-2}$ . 154 systems with $N_{HI}\leq 10^{16}cm^{-2}$
and $N_{CIV}\leq 10^{14}cm^{-2}$ can be related to the
intergalactic ones while 32 systems with $N_{HI}\geq
10^{16}$cm$^{-2}$ are probably linked with intervening galaxies.
The separation of these metal systems can be compared with
results obtained for 268 publicly available metal systems
observed with intermediate resolution.

As is seen from Fig. \ref{fmns_bks}, the redshift distribution of
the metal systems is non homogeneous and the majority of systems
are concentrated at 2~$\leq z\leq$~3.5\,. This means that some of
the discussed here characteristics are derived mainly from this
range of redshifts. At $z\geq~3.5$ the statistics of lines is
not sufficient. Detailed discussion of the observed characteristics
of these metal systems can be found in Boksenberg, Sargent \&
Rough (2003).

\section{Observed characteristics of metal systems}

\begin{figure}
\centering
\epsfxsize=7.5cm
\epsfbox{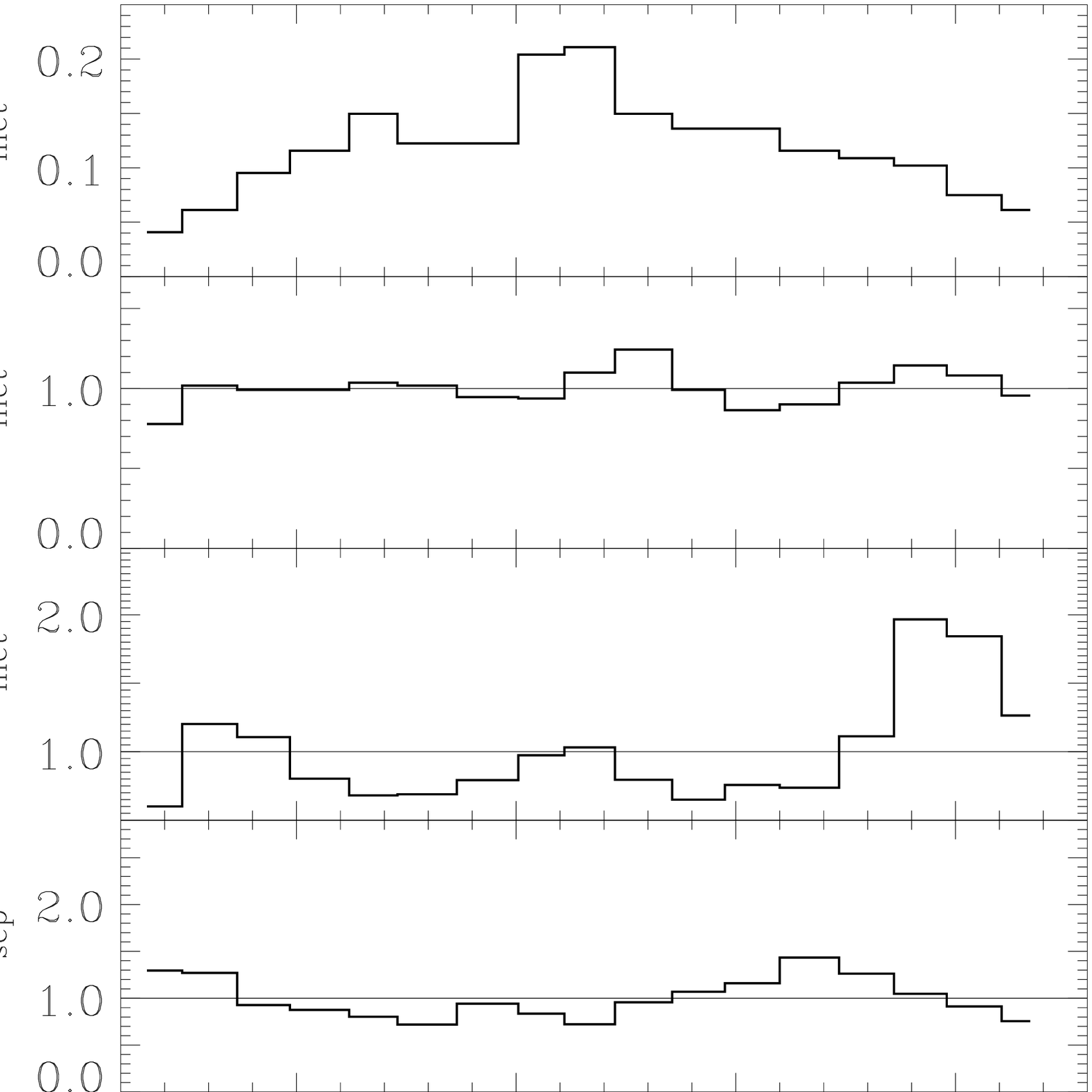}
\vspace{1.cm} 
\caption{For the sample of 154
metal systems the redshift distribution of the metal systems,
$f_{met}$, the mean Doppler parameter, $\langle b_{met}\rangle$, 
the mean carbon abundance in the IGM, $\langle\Omega_{C}^*\rangle
=\langle\Omega_Cz_4^2 \rangle$ \, and the mean system separation, 
$\langle D_{sep}^* \rangle=\langle D_{sep}z_4^{1.75}\rangle$ are 
plotted vs. redshift $z$. }
\label{fmns_bks}
\end{figure}

The sample of 154 metal systems with $N_C\leq 10^{14}cm^{-2}$,
and $N_H\leq 10^{16}cm^{-2}$ can be used to characterize the
metal systems actually related with the IGM. For this sample
the redshift variations of the four mean observed characteristics,
namely, the column density of $CIV$, $\langle N_C\rangle$, the
Doppler parameter, $\langle b_{met}\rangle$, the mean
abundance of carbon in the IGM, $\langle\Omega_C\rangle$, and
the mean separation of metal systems, $\langle D_{sep}\rangle=
\langle D^*_{sep}z_4^{-1.75}\rangle$, $z_4=(1�)/4$ are plotted
in Fig. \ref{fmns_bks} for $2\leq z\leq$ 4 . Here the
mean abundance of the carbon in IGM is estimated as follows,
\[
\langle \Omega_C\rangle =\left\langle\frac{m_CN_C}{D_{sep}}
\frac{8\pi G}{3H_0^2}\right\rangle\,.
\]

The redshift variations of these parameters are fitted by 
expressions
\[
\langle b_{met}\rangle = (11.2\pm 1){\rm km/s}\,,
\]
\[
\langle N_C^*\rangle=\langle N_Cz_4^2\rangle=
3\cdot 10^{13}(1\pm 0.3)cm^{-2},\quad z_4=(1�)/4\,,
\]
\be
\langle\Omega_C^*\rangle=\langle\Omega_Cz_4^2\rangle=
[4\pm 2]h^{-1}\cdot 10^{-8}\sqrt{\Omega_m/0.3}\,,
\label{mns_bks}
\ee
\[
\langle D_{sep}^*\rangle=\langle D_{sep}z_4^{1.75}\rangle =
(27.4\pm 5.8)h^{-1}\sqrt{0.3/\Omega_m}~{\rm Mpc}\,,
\]
respectively. These estimates are consistent with those obtained in
Boksenberg et al. (2003) for 'complex systems'. They are also quite
similar to the estimates of Scannapieco et al. (2006) and Songaila
(2001). For all 193 systems the mean separation decreases to
\be
\langle D_{sep}^*\rangle=\langle D_{sep}z_4^{1.75}\rangle =
(23.8\pm 3.4)h^{-1}\sqrt{0.3/\Omega_m}{\rm Mpc}\,.
\label{allsyz}
\ee

For 32 systems with $N_{HI}\geq 10^{16} cm^{-2}$ which can be
related to intervening galaxies the hydrogen and metals abundance
increase up to:
\be
\langle \lg N_C/N_{HI}\rangle= -4.2\pm 0.6,\quad \langle
\lg N_C\rangle= 14\pm 0.4\,,
\label{nch}
\ee
what indicates the low concentration of carbon as compared with 
its solar abundance $(\lg N_C/N_{HI})_\odot\approx -3.5$ (Allende
Prieto et al. 2002). The mean separation of these objects is
\[
\langle D_{sep}^*\rangle=\langle D_{sep}z_4^{1.75}\rangle =
(93\pm 24)h^{-1}\sqrt{0.3/\Omega_m}{\rm Mpc}\,,
\]
what substantially exceeds the previous ones. For 26 Ly--limit
systems with $N_{HI}\geq 10^{17}cm^{-2}$ this separation increases
up to
\be
\langle D_{sep}^*\rangle=\langle D_{sep}z_4^{1.75}\rangle =(111\pm
31)h^{-1}\sqrt{0.3/\Omega_m}{\rm Mpc}\,,
\label{lylim}
\ee
and the carbon abundance drops down to
\[
\langle \lg N_C/N_{HI}\rangle= -4.6\pm 0.2\,.
\]
For the 268 metal systems selected from spectra observed with
intermediate resolution the mean separation is similar to
(\ref{lylim}).

\begin{figure}
\centering \epsfxsize=7.5cm
\epsfbox{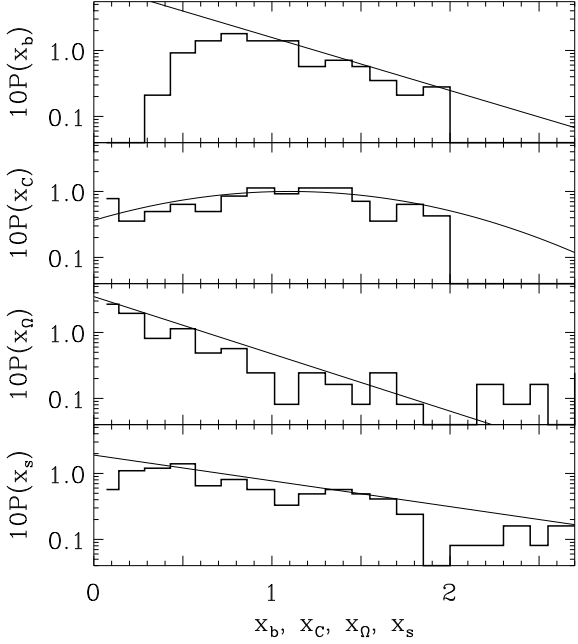}
\vspace{1.cm} 
\caption{For the sample of 193 metal systems PDFs for the 
Doppler parameter, $P(x_b)$, the carbon column density, 
$P(x_C)$, and abundance, $P(x_\Omega)$, and the
separation of metal systems, $P(x_{sep})$, are plotted 
together with fits (\ref{hst_bks}).}
\label{fhst-bks}
\end{figure}

The observed probability distribution functions, PDFs, for the
Doppler parameter, $P(b)$, the carbon column density, $P(
N_{C}z_4^2)$, the carbon abundance, $P(\Omega_C)$, and the
separation of metal systems, $P(D_{sep}z_4^{1.75})$, are plotted
in Fig. \ref{fhst-bks}. Such choice of variables allows us to
suppress the artificial redshift variations of the PDFs caused
by the redshift evolution of the mean characteristics of these
systems. These PDFs are well fitted by the following functions
\[
P_{fit}(x_C)\approx \exp[-(x_C-0.8)^2/1.2],\quad x_C={\lg N_C
/\langle \lg N_C\rangle}\,,
\]
\[
P_{fit}(x_b)\approx \exp(-1.85 x_b),\quad
x_b={b_{met}/\langle b_{met}\rangle}\geq 1\,,
\]
\be
P_{fit}(x_\Omega)\approx 3.5\exp(-2 x_\Omega),\quad x_\Omega=
{\Omega_Cz_4^2/\langle \Omega_Cz_4^2\rangle}\,,
\label{hst_bks}
\ee
\[
P_{fit}(x_{s})\approx 1.9\exp(-x_{s}),\quad x_{s}={D_{sep}
z_4^{1.75}/\langle D_{sep}z_4^{1.75}\rangle}\,,
\]
where again $z_4=(1�)/4$.

For the sample of 193 metal systems the main fraction of observed
systems ($\sim$ 70\%) is linked with rich absorbers with
$\lg(N_{HI})=15\pm 1$ and only 10\% of systems can be related to the
Lyman--limit systems with $\lg(N_{HI})\geq 17$. At the same time,
75\% of absorbers with $\lg(N_{HI})\geq 14$ do not contain metals
above the measured limit $\lg(N_C)\approx 12$.
These results confirm that the observed pollution of the IGM is
caused by rare sources of metals.

It is interesting also to compare the separations of metal
systems with different richness. Thus for subsamples of 91 poorer
systems with $N_c\leq 10^{13}cm^{-2}$ and 102 richer systems with
$N_c\geq 10^{13}cm^{-2}$ the mean separations are quite similar,
$\langle D_{sep}z_4^{1.75}\rangle\sim 32h^{-1}$Mpc. This result
is well consistent with the mean separation obtained for the full
sample of 193 systems (\ref{allsyz}). It emphasises similarity in
the spatial distributions of poorer and richer systems concentrated
within rare clouds.

In spite of limited range of observed redshifts and the large
scatter of estimates (\ref{mns_bks}), growth with time of
$\langle\Omega_C\rangle$ and $\langle N_C\rangle$ points out to
the progressive enrichment of IGM by metals. These results are
consistent with those obtained by Boksenberg et al. (2003) and
Scannapieco (2006) and can be related to the evolution of
galaxies at the same redshifts. However, Songaila (2001) observed
very slow redshift variations of the carbon abundance $\Omega_C
\simeq (5.2\pm 1.7)\cdot 10^{-8}$ at $2\leq z\leq 5.5$ while
Schaye et al. (2003) discussed the slow evolution of unexpectedly
high carbon abundance $\Omega_C\approx 2\cdot 10^{-7}$.

At the same time, the weak redshift variations of both $\langle
N_C\rangle$ and $N_C/N_{HI}$ (\ref{nch}) indicate that for the
stronger systems related to the intervening galaxies the carbon
abundance is redshift independent. The properties of metal
systems are very important for reconstruction of galaxy evolution
at high redshifts and they deserve more detailed observational
investigation.

\section{Discussion}

Results presented in previous Section allow us to test the
degree of relaxation of matter compressed within such absorbers
and to obtain indirect estimates of spatial characteristics of
metal systems. Finally we can estimate the possible contribution
of stars and the process of metal production to the radiation
that reionized  the Universe.

\subsection{Degree of absorbers relaxation}

Comparison of the Doppler parameters of metal systems,
$b_{met}$, and accompanied hydrogen lines, $b_{H}$, allows to
clarify the complex problem of the internal structure of
absorbers. Thus, numerical simulations show that absorbers can
be relaxed along one or two shorter axes and the width of the
absorption lines depends upon the thermal broadening, $b_T$, the
differential Hubble flow, $b_{Hu}$, and turbulent macroscopic
velocities, $b_{mt}$,
\[
b=\sqrt{b_T^2+b_{Hu}^2+b_{mt}^2}\,.
\]
where $b_T\propto m_i^{-1/2}$ depends upon the mass of ion, 
$m_i$, while two other components are the same for all ions.

The relative contribution of these factors varies from absorber
to absorber (see, e.g., Theuns, Schaye \& Haehnelt 2000; Schaye
2001). For majority of absorbers the possible contribution of
Hubble flow is naturally linked with the retained expansion of
clouds and is more essential for weaker absorbers. Its
contribution depends upon the (unknown) relative orientation of
absorber expansion and the line of sight. On the other hand, the
supersonic macroscopic motions are rapidly transformed to the
shock waves and so they disappear. Their life time
 is relatively short and existence over an extended period is
quite problematic what implies that $b\leq\sqrt{2}b_T$ is a more
probable result.

Comparison of the Doppler parameters of metal systems and
accompanied hydrogen lines allows to discriminate between
contributions of thermal and various large scale bulk and
turbulent motions. Indeed, macroscopic and turbulent velocities
are the same for all ions and in that case we can expect that
\be 
b_H\approx b_C,\quad b_{ch}=\sqrt{m_C/m_H}b_C/b_H\sim\sqrt{12}
\approx 3.5\,. 
\label{bch} 
\ee
On the contrary, for any given
temperature of the gas the thermal velocities of ions depend
upon their masses and in this case $b_{ch}$ determined by
(\ref{bch}) should be close to one.

For 93 systems plotted in Fig \ref{ch_bks} comparison of these
Doppler parameters shows that
\be
\langle b_{ch}\rangle=\sqrt{m_C/m_H}\langle b_C/b_H\rangle \approx
1.1\pm 0.5\,,
\label{bch_obs}
\ee
and only for 8 absorbers $2\geq
b_{ch}\geq 1.5$. This result indicates that the thermal
velocities dominate within majority of absorbers in the sample,
$b_T\geq b_{mt}$. As is seen from Fig. \ref{ch_bks} a scatter of
this ratio increases for small $N_C\leq 10^{13}cm^{-2}$ what
indicates smaller degree of relaxation for poorer metal systems.
This scatter increases if the pollution of IGM is caused by the
ejection of metals from closest galaxies when the remaining
velocity of ejection increases the Doppler parameter of metals
and makes it larger than that for the hydrogen component (see,
e.g., Songaila 2006).

These results imply the approximate hydrostatic equilibrium of
compressed matter along the line of sight or at least along the
shorter axis of clouds and agree well with inferences of Carswell,
Schaye \& Kim (2002); Telfer et al. (2002); Simcoe, Sargent \& Rauch
(2002), (2004); Boksenberg, Sargent and Rauch (2003); Bergeron
\& Herbert-Fort (2005). In particular, comparison of the Doppler
parameters measured for HI, CIV and OVI (Carswell, Schaye \& Kim
2002) verifies also that as a rule the macroscopic (turbulent)
velocities are subsonic.

These observational results strongly
suggest that most absorbers are long--lived partly
gravitationally bound and partly relaxed and they are composed
of both DM and baryonic components.

\subsection{Links between metal systems and galaxies}

Natural links between the metal systems and galaxies was
confirmed by direct observations by Adelberger (2005a, b),
 where for majority of systems with
$N_C\sim 10^{14} cm^{-2}$ the proper sizes of absorbing regions
at $z=2 - 3$ were estimated as $R_{abs}\sim 0.08$Mpc for CIV
lines and as $R_{abs}\sim 0.4$Mpc for OVI lines. At the same
time, Scannapieco et al. (2006) found that the metal systems are
clustered within ``bubbles'' with the comoving size $R_{bb}\sim
2h^{-1}$Mpc. Analysis of the observed absorber separations
allows us to test these results.

The mean comoving separation of clouds is determined as 
\be
\langle D_{sep}\rangle =\left\langle\frac{1+z}{3n_{cl}(z)
S_{cl}(z)} \right\rangle\,, 
\label{gsep} 
\ee 
where $n_{cl}\,\&\,S_{cl}$ are the proper number density and 
surface area of the clouds projection on the sky. For the metal 
systems with $N_{HI}\geq 10^{16}cm^{-2}$ we have $\langle 
D_{sep}\rangle\approx 100z_4^{-1.75}h^{-1}{\rm Mpc}$. It can 
be expected that such systems are related to gravitationally
bounded intervening galaxies and for them $S_{cl}$ only weakly
depends on redshift, $S_{cl}\approx const$. Therefore, for such
systems $\langle n_{cl}\rangle\propto (1+z)^{2.75}$ what implies
that at redshifts $z\sim 2-4$ the comoving number density of
galaxies increases with time as $n_{gal}\propto (1+z)^{-0.25}$.
Comparison of the observed number density of galaxies in the
SDSS and 2dF surveys (Doroshkevich et al. 2004) 
\be 
\langle
n_{gal}(z=0)\rangle\sim 10^{-2}h^3 {\rm Mpc}^{-3}\,,
\label{ngal0} 
\ee 
and of LBG galaxies at $z\sim 3$ (Adelberger 2005a), 
\be 
\langle n_{gal}(z=3)\rangle \sim 4h^3\cdot
10^{-3}{\rm Mpc}^{-3}\,, 
\label{ngal3} 
\ee 
indicates similar evolution, $n_{gal}\propto (1+z)^{-1/2}$. 
Using (\ref{lylim}\,\&\,\ref{ngal3}) we get for the proper 
mean surface area and size of such metal systems 
\be 
\langle
S_{cl}\rangle\sim 0.05h^{-2}{\rm Mpc^2},\quad \langle
R_{cl}\rangle=\sqrt{S_{cl}/\pi}\sim 0.12 h^{-1}{\rm Mpc}\,.
\label{sgal} 
\ee 
This estimate is similar to the proper sizes of absorbing 
regions obtained by Adelberger (2005b) at $z=2 - 3$ for CIV 
lines.

For weaker metal systems we can expect progressive growth of the
proper cloud surface area with time, $\langle S_{cl}\rangle
\propto (1+z)^{-\beta}$ and for $\beta\geq 0.25$ a successive
decrease with time of their comoving number density $\langle
n_{cl}\rangle\propto (1+z)^{\beta-0.25}$. This means that due to
clouds expansion some fraction of the weak metal systems goes
under the observational limit $N_C\leq 10^{12}cm^{-2}$. Such
effect was found for the Ly-$\alpha$ forest (Demia\'nski et al.
2006).

Using the estimate (\ref{mns_bks}) for the mean system separation 
we get for the mean proper surface area and size of such systems
\be
\langle S_{cl}\rangle\approx 0.2h^{-2}z_4^{-\beta}Mpc^2, \quad
\langle R_{cl}\rangle\approx 0.25h^{-1}z_4^{-\beta/2}Mpc\,,
\label{fpgal}
\ee
where $z_4=(1+z)/4$. 
This estimate of $R_{cl}$ is of about two -- three times smaller
than the values found by Adelberger et al. (2005b) for weak
lines OVI and the comoving size of ``bubbles'', $R_{bb}\sim
2h^{-1}$Mpc, found by Scannapieco et al. (2006) at redshifts 
$z=2 - 3$.

Formation of galaxies with the baryonic mass $M_b$ implies
accumulation of matter from the protogalaxy with the proper size
\be 
R_{prg}\approx (6V/\pi)^{1/3}\approx 0.4z_4^{-1}Mpc
\left[\frac{M_b}{10^{10}M_\odot}\frac{0.02}{\Omega_bh^2}
\right]^{1/3}\,, 
\label{sizeprg} 
\ee 
what quite well agrees with (\ref{fpgal} for $M_b\sim 10^9 
M_\odot$ and with observations of Adelberger et al. (2005b) 
and Scannapieco et al. (2006) for $M_b\sim 10^{10}M_\odot$.

Now the pollution of IGM by metals is related either to ejection
of metals from galaxies after explosions of supernovae or to the
stripping of satellites of the central galaxy. It is most
probable that both processes are equally important. Thus, as was
noted in Sec. 3.2, the complex structure of richer systems
implies the multistep metal ejection what is naturally connected
with SN explosions in the large central galaxy. On the contrary,
at large distances from the central galaxy ejection of metals
from satellites and their stripping dominate.

The pronounced correlation between the observed galaxies and
metal systems reflects the expected interaction between the
large and small scale perturbation when former ones modulate the
process of galaxy formation and amplify the inhomogeneities in
the spatial galaxy distribution. At the same time, the noted
above similarity in the spatial distributions of richer and
poorer metal systems indicates similarities in their genesis and
their close links with the ordinar observed galaxies.

\subsection{Probable sources of reionization}

Measurements of the carbon abundance in the IGM at high
redshifts allows us to clarify some aspects of the process of
reionization of the Universe. Thus, the concentration of metals
in the older galaxies observed at redshifts $z\sim 6 -7$ and
higher is the integral measure of the contribution of nuclear
reactions within any stars to the creation of the ionizing UV
radiation. On the other hand, the comparison of the carbon
abundances within galaxies and in the IGM characterizes the
efficiency of the processes of star explosion and ejection of
metals from the host galaxies and their dwarf satellites. It is
important that this approach estimates the integral action of
all stellar sourses ignoring the often discussed (see, e.g.
Choudhury et al. 2007) but badly known details of the
reionization process such as the mass function of the first
galaxies and Pop. III stars, the rate of the star creation,
radiative feedback or the redshift variations of the transmitted
flux and the electronic depth.

\subsubsection{Contribution of observed galaxies}

As is well known, the transformation of hydrogen to carbon
produces $\approx 7.3MeV$ of energy per one baryon what is
equivalent to $N_\gamma\approx 5\cdot 10^5$ of UV photons with
energy $E_\gamma \approx 13.6 eV$. The synthesis of $O, N, Si$
and other metals produces the UV radiation with the similar
efficiency. According to more detailed estimates (see, e.g.,
Tumlinson et al. 2004) the value of ionizing photons per one
baryon is smaller and varies with the mass of Pop. III stars
in the range $N_\gamma\simeq (2 -8)\cdot 10^4$.

At the same time, the observed strong ionization of hydrogen at
$z= 5-7$ implies the generation already at such redshifts of at
least one UV photon per baryon ($N_{b\gamma}\geq 1)$. Of course,
this is the minimal value and in some papers (see, e.g.
Dijkstra, Haiman, Loeb 2004; Madau 2007) production of extra (up
to 10) UV photons per baryon is discussed. Thus, if the
reionization was caused by the production of metals within stars
then the minimal abundance of metals at $z\sim 6$ must be at
least
\be
\Omega_{min}\simeq\frac{\Omega_bN_{b\gamma}}{f_{esc}N_\gamma}
\approx 0.8\cdot 10^{-7}\frac{N_{b\gamma}}{f_{esc}}\frac{5\cdot 
10^5}{N_\gamma}\frac{\Omega_b}{0.04}\,,
\label{min}
\ee
where $N_{b\gamma}\geq 1$ is the minimal number of UV photons per
one baryon required for the reionization and $f_{esc}\leq 1$ is
the mean fraction of UV photons escaping from galaxies. The
allowance for the complex spectral distribution of the generated
UV photons, ionization of He and the heating and slow
recombination of the IGM increases this estimate of the
$\Omega_{min}$ by a factor of 2 -- 3.

\begin{table}
\caption{Galaxies at $z\geq 3$}
\label{tbl1}
\begin{tabular}{cccl}
$\langle z\rangle$&$M_*$&$\rho_*$&Ref. \\
              &10$^9M_\odot$&10$^6M_\odot${\rm Mpc}$^{-3}$&  \\
$\sim$ 3&  0.1&0.3  & Lai et al. (2007)\\
$\sim$ 3&   1 &3.   & Lai et al. (2007)\\
$\sim$ 5&   1 &1.5  & Verma et al. (2007)\\
$\sim$ 6&   1 &1 - 7& Yan et al. (2006)\\
$\sim$ 6&   10&2.5  & Eylis et al. (2007)\\
$\sim$5-6&50 - 500&8.&Wiklind  et al. (2007)\\
$\sim$ 7& 1-10&1.6  & Labbe et al. (2006)\\
\vspace{0.15cm}
\end{tabular}

$M_*$ and $\rho_*$ are the mean stellar mass and the stellar 
mass density for the observed sample of galaxies.
\end{table}

Available now estimates of the stellar component in galaxy
populations observed at redshifts $z\geq 3$ are summarized in
Table 1\,. Here observations of Lai et al. (2007) relate to the
two populations of LAEs (younger and older) and the scatter of
measured $\rho_*$ is determined by variations of the selected
samples of galaxies.

The most interesting estimates for our goals came from recent
observations of 11 galaxies by Wiklind et al. (2007) (Table
\ref{tbl1}) which are related to older massive compact galaxies
with the comoving number density $n_{gal}\approx 1.4\cdot
10^{-5} {\rm Mpc}^{-3}$, the proper sizes $R_g\sim$2kpc and ages
0.2--1 Gyr. It is expected that the majority of these stars were
formed at $z\geq 9$ and they represent the objects actually
responsible for the reionization.

For this sample of galaxies we get for the stars and
metal abundances
\be
\Omega_*\approx 6.7\cdot 10^{-5},\quad \Omega_{*met}=\Omega_*Z
\approx 1.3\cdot 10^{-7}{Z\over 0.1Z_\odot}\,,
\label{met-5}
\ee
and $\Omega_*$ is significantly below the present day values
$\Omega_{lum}(z=0)\approx 4\cdot 10^{-3}$ (Fukugita et al. 1998).
It is evident that this result is consistent with $\Omega_{min}$
(\ref{min}) only for an unlikely value $N_{b\gamma}/f_{esc}\sim 1$.

On the other hand, any such galaxy can ionize baryons in a
volume with the mass $M_i$ and comoving size $R_i$ where
\be
M_i\sim 20\epsilon_{eff}M_*,\quad \epsilon_{eff}=\frac{f_{esc}}
{0.1}\frac{Z}{0.1Z_\odot}\frac{N_\gamma}{5\cdot 10^5}\,,
\label{MRi}
\ee
\[
R_i=\left(\frac{3M_i}{4\pi\langle\rho_b\rangle}\right)^{1/3}\sim
5.5{\rm Mpc}\left(\epsilon_{eff}{M_*\over 10^{11}
M_\odot}\frac{0.02}{\Omega_bh^2}\right)^{1/3}\,.
\]
So the moderate size of ionized bubble corresponds to the small
fraction of the ionized IGM, 
\be
f_i=M_in_{gal}/\langle\rho_b\rangle\sim 10^{-2}\epsilon_{eff}
{M_*\over 10^{11}M_\odot}\frac{0.02}{\Omega_bh^2}\,, 
\label{frac} 
\ee 
what
illustrates the limited efficiency of ionization caused by the
nuclear reactions within such galaxies. Of course, as was noted
above, at redshift $z=3$ the density of LBG galaxies
(\ref{ngal3}) is larger by a factor of $10^2$ but for them
$M_*\ll 10^{11}M_\odot$ (Table \ref{tbl1}) and finally we
get again $f_i\ll 1$.

Estimates of the star radiation (\ref{min}, \ref{frac}) include
the factor $f_{esc}$. Its available estimates are quite
uncertain as they strongly depend upon the internal structure of
galaxy and the immediate environment of the UV sources. These
estimates vary from $f_{esc}\sim$2 -- 3\% to $f_{esc}\sim$10\%
(Iliev et al. 2004; Dijkstra et al. 2004; Meiksin 2005). For
example, for the observed LAEs $f_{esc}\rightarrow 0$ and so
majority of the UV radiation is absorbed within the galaxy.
Thus, we can conclude that either all observations cited above
and Eq. (\ref{met-5}) strongly underestimate the metal
production and the UV radiation of high redshifts galaxies or
the reionization is related with some unobserved thermal and/or
non thermal sources of UV radiation.

\subsubsection{Possible contribution of dwarf galaxies}

The differences between estimates (\ref{min}) and (\ref{met-5})
will decrease if the majority of ionizing UV radiation is
related with the dwarf unobserved galaxies. Such preferential
formation of dwarf galaxies is natural in the CDM cosmology and
is consistent with observations (see, e.g., Yan \& Windhorst
2004; Yan et al. 2006; Stark et al. 2007). It is also consistent
with the shape of the UV luminosity function observed at
redshifts $z=4 - 6$ (Bouwens et al. 2007). Moreover,
reconstruction of the initial power spectrum in Demia\'nski et
al. (2006) indicates the possible excess of power at scales
$M\leq 10^8 M_\odot$ what can additionally stimulate earlier
formation of dwarf galaxies.

The contribution of dwarf galaxies is also restricted owing to
the observational limitations for the small fraction of dwarf
galaxies survived at $z=0$ (see also Adelberger 2005a). Non the
less, the important role of dwarf galaxies in the reionization
is supported by the observed abundance of CIV in the IGM. All
published estimates of carbon abundance at redshifts $3\leq
z\leq 5.5$ are close to $\Omega_C\approx (5.2\pm 1.7)\cdot 
10^{-8}$ (Songaila 2001; Pettini et al. 2003; Boksenberg et al. 
2003; Scannapieco et al. 2006; Ryan-Weber et al. 2007) and are 
consistent with (\ref{mns_bks}) in the range of scatter. The 
slow redshift evolution of the abundance implies that the main 
enrichment of the IGM takes place at high redshifts $z\geq 6$. 
Moreover, the carbon abundance in the IGM is also close to that 
observed in galaxies at the same redshifts (\ref{met-5})
\be
\Omega_{*C}=\Omega_{*met}Z_C/Z_{met}\simeq 0.2\Omega_{*met}
\simeq 2.6\cdot 10^{-8}\,.
\label{civ-5}
\ee

As was noted in the previous subsection these metals are
presumably concentrated within bubbles with the comoving size
$\sim 2h^{-1}$ Mpc. This size significantly exceeds the typical
size $\sim 2$ kpc observed for galaxies at the same redshifts
(Wiklind et al. 2007) but it is close to the size of protogalaxy
with the baryonic mass $M_b\sim 10^{10} - 10^{11}M_\odot$. In
turn, such patchy spatial distribution of metals in the IGM
implies that the process of metal enrichment is dominated by
metal ejection from the dwarf satellites of the host galaxy in
the course of their hierarchical clustering. This inference is
close to that formulated in Songaila (2006) where it was noted
that the formation of weak CIV systems cannot be easily
understood in terms of high velocity galactic wind. Observation
of GRB at redshift $z=6.3$ (Haislip et al. 2005) can be
considered as a possible example of such processes. This
discussion shows that the real metal abundance at $z\geq 6$
exceeds the estimate (\ref{met-5}) at least by a factor of 2.
On the other hand, such scenario implies also the existence at
$z=0$ of the invisible dwarf galaxies in the vicinity of massive
galaxies of early types.

\subsubsection{Possible contribution of nonthermal sources}

This discussion shows that even taking into account  all
uncertainties of observations the attempts to explain the
reionization as a consequence of nuclear reactions within stars
seem to be highly problematic (see also Adelberger 2005a;
Meiksin 2005). The non thermal sources of UV photons such as
AGNs and black holes formed both in pregalactic systems and
within first galaxies produce $\sim 50$Mev per one accreted
baryon what is 7 times more effective than the nuclear
reactions. This means that even accretion of a small fraction of
baryons, 
\be 
\Omega_{acr}\simeq 10^{-8}/f_{esc}\simeq 1.4\cdot
10^{-4} \Omega_*/f_{esc}\,, 
\label{accret} 
\ee 
will produce the
same effect as the observed stars (\ref{met-5}). This verifies
that such non thermal sources of UV radiation can be considered
as very promising ones and that they can be actually responsible
for the reionization (see, e.g., Madau \,\&\,Rees 2001; Meiksin
2005; Reed 2005; Ciardi et al. 2006; Madau 2007).

The observational restrictions of such non thermal sources of
the UV radiation are poor. Thus, Meiksin notes that the comoving
space density of galaxies with the black holes is $n_{bh}\sim
2\cdot 10^{-4}$Mpc$^{-3}$ what is $\sim 1\%$ of galaxies at
$z=0$. However, this value is 10 times more than the comoving
number density of early massive galaxies (Wiklind et al. 2007)
which demonstrate some features of the AGN activity. A faint AGN
was directly observed at $z=5.44$ (Douglas et al. 2007).
Restrictions of the contribution of such sources discussed in
Dijkstra et al. (2004) were criticized by Meiksin (2005). The
possible contribution of exotic sources such as the unstable
fraction of DM particles or antimatter can be noticeable at
$z\geq 20 - 30$ (Cohen et al. 1998; Bambi \& Dolgov 2007; Freez
at al. 2007) but it cannot dominate at the period of
recombination. Its possible impact will be soon tested by the
Planck mission.

These problems are open for discussions but the important
contribution of non thermal UV sources seems to be required.
Further progress can be achieved with a richer and more refined
sample of observed metal systems and earlier galaxies.

\subsection*{Acknowledgments}
This paper was supported in
part by the Polish State Committee for Scientific Research grant Nr.
1-P03D-014-26 and Russian Found of Fundamental Investigations grant
Nr. 05-02-16302.


\begin{thebibliography}{}

\bibitem[2005]{adelberg}
Adelberger K., 2005a, Proceeding IAU Colloquium No.199,
``Probing galaxies throwgh Quasar Absorption lines'',
IAU Colloquivium No.199, eds. P.Williams, C.Shu, and B.Menard,
astro-ph/0504311

\bibitem[2005]{adelberg}
Adelberger K., Shapley A., Steidel C., Pettini M., Erb D.,\,\&\,
Reddy N., 2005b, ApJ, 629, 636

\bibitem[2002]{Allende}
Allende Prieto C., Lambert D., Asplund M., 2002, ApJ, 573, L137

\bibitem[2007]{BD}
Bambi C., Dolgov A., 2007, astro-ph/0702350

\bibitem[2001]{Becker-et-al}
Becker~R.H. et al. 2001, AJ, 122, 2850

\bibitem[1992]{Bergeron etal}
Bergeron~J., Cristiani~S., \& Shaver~P.A., 1992, A\&A, 257, 417

\bibitem[1992]{Bergeron -H}
Bergeron~J. \& Herbert-Fort~S., 2005, ``Probing Galaxies
through Quasar Absorption lines'', IAU Colloquivium No.199,
eds. Williams~P.R., Shu~C., \& Menard~B.. astro-ph/0506700

\bibitem[2007]{Bouwens}
Bouwens R., Illingworth G., Franx M., Ford H., 2007,
arXiv:astro-ph/0707.2080v2

\bibitem[2003]{Boksenberg-et-al}
Boksenberg A., Sargent W.L.W., Rauch M., 2003, ApJS, submit.
astro-ph/0307557

\bibitem[2003]{Carsw}
Carswell~R., Shaye~J., \& Kim~T.S., 2002, ApJ, 578, 43

\bibitem[2007]{Choudry}
Choudry R., Ferrara A., Gallerani S., astro-ph/0712.0738

\bibitem[2006]{Ciardi-et-al}
Ciardi B., Scannapieco E., Stoehr F., Ferrara A., Iliev I.,
Shapiro P., 2006, MNRAS, 366, 689

\bibitem[1998]{Cohen-et-al}
Cohen A., De Rujula A., Glashow S., 1998, ApJ, 495, 539

\bibitem[2004]{dij}
Dijkstra M., Haiman Z., Loeb A., 2004, ApJ, 613, 646

\bibitem[2001]{djor}
Djorgovski S., Castro S., Stern D., Mahabal A., 2001,
ApJ, 560, L5

\bibitem[2006]{dem-et-al}
Demia\'nski M., Doroshkevich A., Turchaninov V., 2006, MNRAS,

\bibitem[2004]{dorr-et-al}
Doroshkevich A., Tucker D.L., Allam S., Way M.J., 2004, AA, 418, 7

\bibitem[2007]{douglas}
Douglas L., Bremer M., Stanway E., Lehnert M., 2007, MNRAS, 376,
1393

\bibitem[2007]{Eyles}
Eyles L., Bunker A., Ellis R., Lacy M., Stanway E., Stark D.,
Chiu K., 2007, MNRAS, 374, 910

\bibitem[2004]{Fan-et-al}
Fan~X. et al., 2004, AJ, 128, 515

\bibitem[2007]{Freez}
Freese K., Gondolo P., Spolyar D., 2007, astro-ph/0709.2369

\bibitem[2002]{Frye}
Frye B., Broadhurst T., Benitez N., 2002, ApJ, 568, 558

\bibitem[1998]{Fukugita-et-al}
Fukugita M., Hogan C., Peebles P.J.E., 1998, ApJ, 503, 518


\bibitem[2006]{Hinshaw}
Hinshaw G. et al., 2007, ApJS, 170, 288

\bibitem[2005]{haislip}
Haislip J., et al., 2006, Nature, 440, 181

\bibitem[2004]{Iliev}
Iliev I, Shapiro P., Raga A., 2005, MNRAS, 361, 405

\bibitem[2006]{labbe}
Labbe I., Bouwence R., Illingworth G., Franx M., 2006, ApJ,
649, L67

\bibitem[2006]{Lai}
Lai K. et al., 2007, arXiv:astro-ph/0710.3384v1

\bibitem[1995]{Lanzetta-et-al}
Lanzetta~K.M., Bowen~D.V., Tytler~D., \& Webb~J.K., 1995,
ApJ.,  442, 538.

\bibitem[1996]{LeBrune-et-al}
Le Brune~V., Bergeron~J., \& Boisse~P., 1996, A\&A,  306, 691


\bibitem[2001]{Madau-Rees}
Madau P., Rees M., 2001, ApJ., 555, 9

\bibitem[2007]{Madau}
Madau P., 2007, ``The Emission Line Universe'', eds. J. Cepa
and N. Sanchez, Cambridge Unuversity Press, astro-ph/0706.0123

\bibitem[2005]{Meiksin}
Meiksin A., 2005, MNRAS, 356, 596

\bibitem[1981]{Oort-81}
Oort~J.H., 1981, Astr.Astrophys.,  94, 359.

\bibitem[1984]{Oort-84}
Oort~J.H., 1984, Astr.Astrophys.,  139, 211.

\bibitem[2001]{pet}
Pettini M., Shapley A., Steidel C., Cuby J., Dickinson M.,
Moorwood A., Adelberger K., Giavalisco M., 2001, ApJ, 554, 981

\bibitem[2003]{petttini}
Pettini M., Madau P., Bolte M., Prochska J., Ellison S., Fan X.,
2003, ApJ, 594, 695

\bibitem[2005]{Reed 2005}
Reed D., Bower R., Frenk C., Gao L., Jenkins A., Theuns T.,
White S., 2005, MNRAS, 363, 393.

\bibitem[2004]{Ricotti}
Ricotti M., Ostriker J.P., 2004, MNRAS, 350, 539

\bibitem[2007]{Ryan 2007}
Ryan-Weber E., Pettini M., Madau P. 2006, MNRAS, 371, L78

\bibitem[1980]{Sargent-et-al}
Sargent,~W.L.W., Young,~P.J., Boksenberg,~A. \&
Tytler,~D.,  1980, Ap.J.Suppl,  42, 41

\bibitem[2005]{scann}
Scannapieco E., Pichon C., Aracil B., Petitjean P., Thacker R.,
Pogosian D., Bergeron J., Couchman H., 2006, MNRAS, 365, 615;
astro-ph/0503001

\bibitem[2007]{Schaerer}
Schaerer D., 2007, ``The Emission Line Universe'', eds. J. Cepa
and N. Sanchez, Cambridge Unuversity Press, astro-ph/0706.0139

\bibitem[2001]{Schaye 2001}
Schaye~J., 2001, ApJ, 559, 507

\bibitem[2003]{schay}
Schaye J., Aguirre A., Kim T.S., Theuns T., Rauch M.,
Sargent W.L.W., 2003, ApJ., 596, 768


\bibitem[2002]{Schneider}
Schneider R., Ferrara A., Natarajan P., Omukai K., 2002, ApJ,
571, 30

\bibitem[2007]{Shull}
Shull M., Venkatesan A., 2007, arXiv:astro-ph/0702323

\bibitem[2002]{Simcoe}
Simcoe~R., Sargent~W.L.W., Rauch~M., 2002, ApJ., 578, 737

\bibitem[2004]{Simco}
Simcoe~R., Sargent~W.L.W., Rauch~M., 2004, ApJ., 606, 92

\bibitem[2007]{smith}
Smith B., Sigurdsson S., 2007, ApJ, 661, L5

\bibitem[2001]{songaila}
Songaila A., 2001, ApJ., 561, L153

\bibitem[2006]{songaila6}
Songaila A., 2006, AJ., 131, 24

\bibitem[2006]{stark}
Stark D., Ellis R., Richard J., Kneib J., Smith G., Santos M.,
2007, ApJ, 663, 10

\bibitem[2004]{Telfer}
Telfer~R.C., Kriss~G.A., Zheng~W., Davidsen~A.F. \& Tytler~D.,
2002, ApJ, 579, 500

\bibitem[2000]{Teuns-et-all.}
Theuns~T., Schaye~J., Haehnelt~M., 2000, MNRAS, 315, 600

\bibitem[2007]{Tornatore.}
Tornatore L., Ferrara A., Schneider R., 2007,
arXiv:astro-ph/0797.1433

\bibitem[2004]{Tumlinson}
Tumlinson J., Venkatesan A., Shull M., 2004, ApJ, 612, 602

\bibitem[1995]{Tytler-95}
Tytler~D., 1995, in Meylan~J., ed., QSO Absorption Lines,
p. 289.

\bibitem[2007]{Verma}
Verma A., Lehnert M., F\"orster Schneider N., Bremer M.,
Douglas L., 2007, MNRAS, 377, 1024

\bibitem[2007]{Wiklind}
Wiklind T., Dickinson M., Ferguson H., Giavalisco M., Mobasher B.,
Grogin N., Panagia N., 2007, arXiv:astro-ph/0710.0406


\bibitem[2004]{YanW}
Yan H., Windhorst R., 2004, ApJ, 600, L1

\bibitem[2006]{Yan}
Yan H., Dickinson M., Giavalisco M., Stern D., Eisenhardt P.,
Ferguson H., 2006, ApJ, 651, 24

\end{thebibliography}
\end{document}